%% file: main.tex
\documentclass[acmtog, natbib=true, screen=true]{acmart}

\usepackage{subfigure}
\usepackage{wrapfig}
\usepackage{adjustbox}
\usepackage{soul}
\usepackage{xspace}
\setcitestyle{square}

\newcommand{\duygu}[1]{\textcolor[rgb]{0.2,0.2,0.5}{\textbf{Duygu: #1}}}

\newcommand{\revision}[1]{#1}

\newcommand{\curveF}{\textsc{CurveFusion}\xspace}
\newcommand{\Lone}{L$_1\xspace$}

\newcommand{\tabincell}[2]{\begin{tabular}{@{}#1@{}}#2\end{tabular}}

\begin{document}
\title
[\curveF: Reconstructing Thin Structures from RGBD Sequences]
{\curveF: Reconstructing Thin Structures from RGBD Sequences}

\setcopyright{acmcopyright}
\acmJournal{TOG}
\acmYear{2018}\acmVolume{37}\acmNumber{6}\acmArticle{218}\acmMonth{11} \acmDOI{10.1145/3272127.3275097}

\author{Lingjie Liu}
\authornote{These two authors contributed equally}
\affiliation{
    \institution{University of Hong Kong}
}
\affiliation{
    \institution{University College London}
}
\email{liulingjie0206@gmail.com}

\author{Nenglun Chen}
\authornotemark[1]
\affiliation{
    \institution{University of Hong Kong}
}
\email{chennenglun@gmail.com}

\author{Duygu Ceylan}
\affiliation{
    \institution{Adobe Research}
}
\email{ceylan@adobe.com}

\author{Christian Theobalt}
\affiliation{
    \institution{Max Planck Institute for Informatics}
}
\email{theobalt@mpi-inf.mpg.de}

\author{Wenping Wang}
\affiliation{
    \institution{University of Hong Kong}
}
\email{wenping@cs.hku.hk}

\author{Niloy J. Mitra}
\affiliation{
    \institution{University College London}
}
\email{n.mitra@cs.ucl.ac.uk}

\renewcommand{\shortauthors}{Liu, Chen, et al.}

\begin{teaserfigure}
  \includegraphics[width=\textwidth]{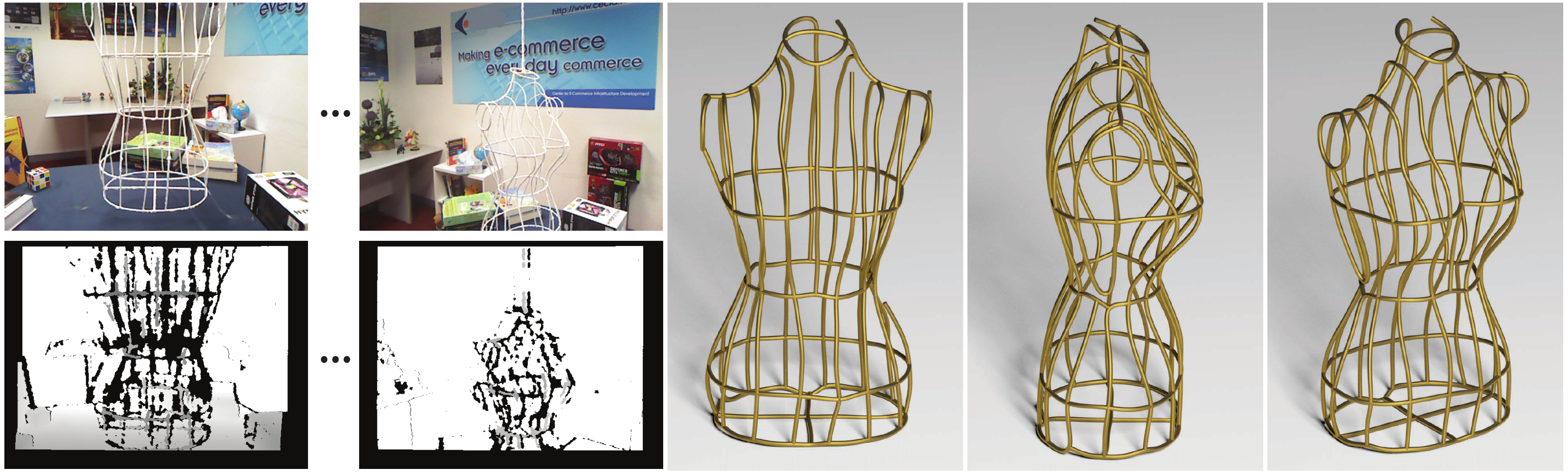} 
  \caption{We introduce \curveF, a method to reconstruct objects made of thin filament-like structures from an RGBD sequence by fusing information from noisy depth scans and loosely using RGB information for verification only.  }
  \label{fig:teaser}
\end{teaserfigure}

\input{abstract}

%
%



\maketitle

\input{introduction_ct}
\input{relatedWork}

\input{overview.tex}
\input{method}

\input{results}

\input{conclusion}
\begin{acks}
We thank our reviewers for their invaluable comments. We thank Hui Huang, Amy Tabb and Zheng Wang for their great help with the testing and validation of our work. We also thank Jiatao Gu, Cheng Lin, Zhiming Cui, Runnan Chen, Maria Lam, Paul Guerrero, Elizabeth Schildge for their help. This work was partially funded by the ERC Starting Grant SmartGeometry (StG-2013-335373), the Research Grant Council of Hong Kong (GRF 17210718), the ERC Consolidator Grant 4DReply (770784), a Google Faculty award, a UCL visiting student program, and gifts from Adobe.
\end{acks}

\bibliographystyle{ACM-Reference-Format}
\bibliography{references}

\end{document}

%% file: abstract.tex
\begin{abstract}

%
%

We introduce \curveF, the first approach for high quality scanning of thin structures at interactive rates using a handheld RGBD camera. 
Thin filament-like structures are mathematically just 1D curves embedded in $\mathbb{R}^3$, and integration-based reconstruction works best when depth sequences (from the thin structure parts) are fused using the object's (unknown) curve skeleton. 
%
%
Thus, using the complementary but noisy color and depth channels, \curveF\ first automatically identifies point samples on potential thin structures and groups them into \emph{bundles}, each being a group of a fixed number of aligned consecutive frames. Then, the algorithm extracts per-bundle skeleton curves using \Lone\ axes, and aligns and iteratively merges the \Lone\ segments from all the bundles to form the final complete curve skeleton. Thus, unlike previous methods, reconstruction happens via integration along a \textit{data-dependent fusion primitive}, i.e., the extracted curve skeleton. 
We extensively evaluate \curveF\ on a range of challenging examples, different scanner and calibration settings, and present high fidelity thin structure reconstructions previously just not possible from raw RGBD sequences.  

\end{abstract}

%% file: introduction_ct.tex
\section{Introduction}

The past few years have seen significant research progress in the context of scanning of large-scale 3D environments. With easy access to commodity depth cameras producing real-time feeds of RGBD sequences, the most successful approach aligns and integrates many such low-quality input frames and extracts a {\em fused} surface.  Well-known examples include KinectFusion~\cite{Newcombe:2011}, BundleFusion~\cite{Dai:2017}, etc. 

The above family of methods implicitly assumes that the scanned environments consist of volumetric objects with closed or extended boundary surfaces. Hence, multiple depth scans, once aligned, can be effectively integrated using a (truncated) signed distance field~(TSDF) representation~\cite{Curless:1996} discretized on a pre-defined voxel grid, and a surface is extracted using Marching Cubes or its variants. 

\begin{figure}[t!]
    \centering
\includegraphics[width=\columnwidth]{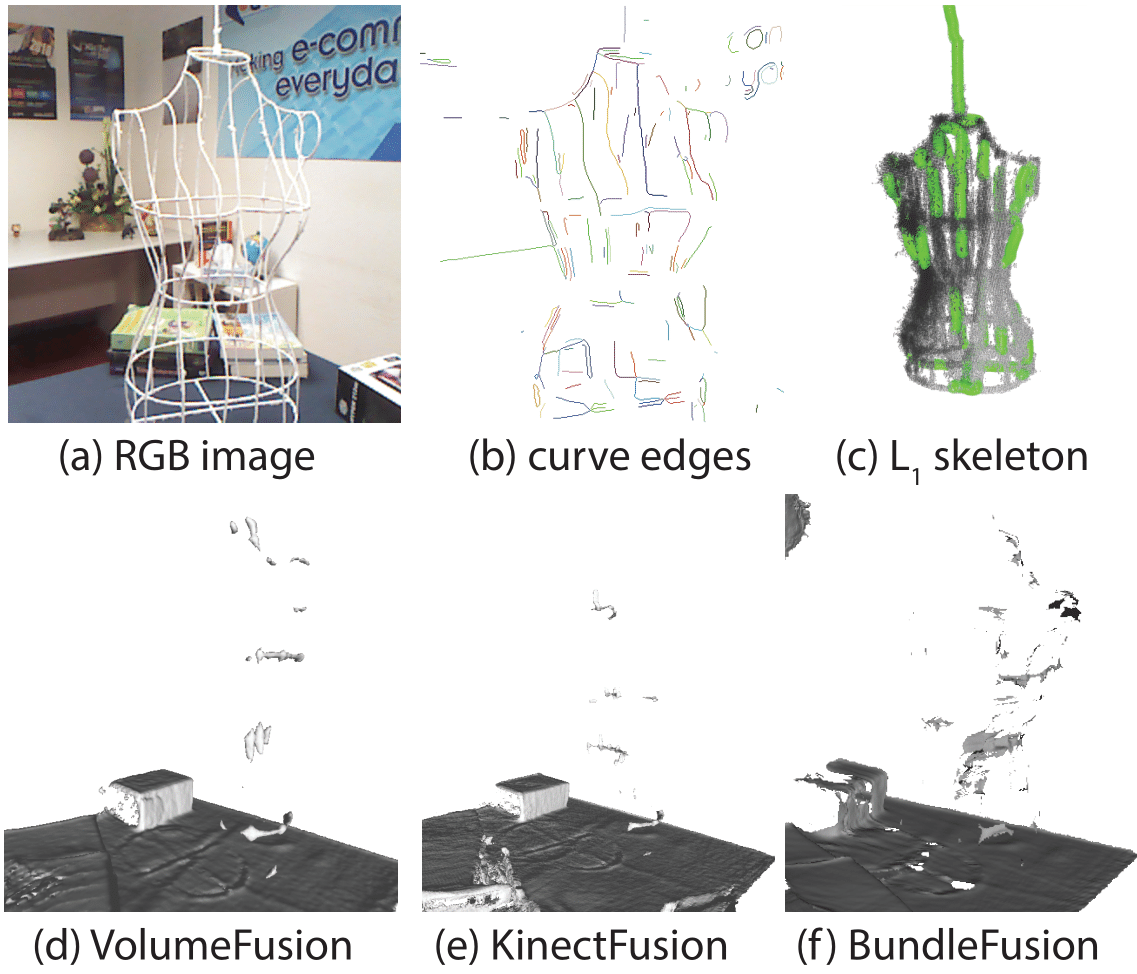}
    \caption{For the model in Figure~\ref{fig:teaser}, voxel based fusion methods fail to recover the thin structure: results shown using VolumeFusion~\protect\cite{Curless:1996}, KinectFusion~\protect\cite{Newcombe:2011}, and  BundleFusion~\protect\cite{Dai:2017}. 
    We show an RGB frame for reference, curves extracted from RGB, and the effect of extracting the L$_1$ axis from the aligned depth data of the whole RGBD sequence, which yields poor reconstruction. }
    \label{fig:RGBD_motivation}
\end{figure}

Such volumetric fusion approaches simply fail to capture objects or object parts primarily consisting of thin or filamentary structures (e.g., features less than $4$~mm in diameter). Figure~\ref{fig:RGBD_motivation} shows  typical results of a scanning session using different fusion approaches: volumetric TSDF~\cite{Curless:1996}, KinectFusion~\cite{Newcombe:2011}, and BundleFusion~\cite{Dai:2017}. The thin parts are either noisy or completely missed in the fused output.

Specifically, capturing thin structures by `fusing' several RGBD frames is challenging for multiple reasons:
(i)~An RGBD camera's depth data is noisy and of limited spatial resolution -- parts of thin structures are regularly missed, or captured at incorrect depths in the raw depth images~\cite{teichman2013unsupervised}, and there is no connectivity information among the few detected points. Combining aligned depth images from multiple viewpoints can only partially recover thin structure data already missed in individual input frames. Further, the inevitable camera drift is particularly detrimental when trying to align thin structure data.  
(ii)~A depth camera's RGB channel often provides complementary information at higher pixel resolution. 
Unfortunately, RGB images of thin structures are equally challenging to detect and segment since it is very difficult to distinguish curve edges from  texture/background edges (see Figure~\ref{fig:RGBD_motivation}b). 
(iii)~Moreover, despite the considerable success of existing works on image segmentation ~\cite{fu2017object,pan2017superpixels,cao2017extracting}, it remains a challenge to segment noisy and partially missing depth images containing thin structures.
Therefore, attempts to recover 3D thin structures, e.g., by lifting 2D curves from RGB images to 3D thin structures using depth images, fail on two counts: first, many incorrect edges are lifted, and unreliable or missing depth measurements prohibit correct lifting. Second, existing methods that use the color channel to upsample the pixel resolution of the depth channel, e.g., through joint filtering~\cite{Kopf:2007:JBU,Richardt:2012:RGBD} or shading-based refinement~\cite{Wu:2014:RSR}, cannot recover raw thin structure depth measurements that are missing or inaccurate to start with. 
%
(iii)~Previously used predetermined fusion data structures (e.g., a regular TSDF voxel grid), being oblivious of the object being scanned, are unsuited for thin scene structures. In this work, we show that central skeletons can be used as {\em adaptive} fusion primitives to encode connectivity and geometry, and therefore can represent thin structures more accurately and more compactly, which greatly simplifies structural completion of missing parts. 

%


We therefore propose {\curveF}, a reconstruction approach for thin structures with a handheld commodity RGBD camera. Instead of using a pre-authored fusion primitive, such as a voxel grid, {\curveF} uses the curve skeleton as a new, data-dependent fusion primitive. 
\curveF~first identifies and segments depth samples that come from thin-structures in each frame using both depth and color information.
Algorithmically, to address the issue that each single raw depth frame is noisy and incomplete, we partition the input frame sequence into {\emph{bundles}}, each being a group of a fixed number of consecutive frames. We further consolidate the segmented point samples from the mutually-aligned frames of the same bundle into a point set, called the {\em bundle difference set}. For each bundle we extract the central skeleton curve of this merged point set by computing its {\Lone} axis~\cite{Huang:2013} and explicitly detect junctions, i.e. where three or more curve segments meet.

\revision{Given many such {\Lone} curve fragments along with identified junctions} extracted from different bundles, we propose a novel curve alignment algorithm to fuse them into the final 3D skeleton curves with minimal drift. 
We retain the junctions through this fusion stage. The extracted 3D skeleton is then used to recover the final 3D surface of the scanned thin structure (see Figure~\ref{fig:teaser}).

We evaluated \curveF\ on a range of RGBD scans of real thin structured objects \revision{under various scanning setups} and demonstrate high-quality reconstructions. We compare our method to competing image-based methods, which either require significantly more controlled imaging setups or start to degrade in complex examples even when assuming access to clean background subtracted inputs. 
Please note that state-of-the-art voxel-based fusion methods simply fail to produce any relevant output for most of the examples shown in this paper. 
In summary,  we (i)~develop the first method for performing high-quality 3D curve reconstruction using commodity RGBD cameras; (ii)~propose a data-adaptive fusion approach that first discovers a suitable skeleton directly from the data and then use it for reconstruction; and (iii)~present high-quality reconstructions of thin structures previously not possible from raw RGBD sequences. 
%

%% file: relatedWork.tex

\section{Related Work}
Our work is related to depth-based reconstruction techniques as well as methods of reconstructing thin structures.

\paragraph*{Depth based reconstruction} The many recent advances in 3D acquisition technology (e.g., structured light, LiDAR, and more recently commodity depth sensors) spawned an increasing number of related works on digitization of the physical world from depth measurements. Since many depth sensors output 3D points, earlier approaches propose to merge these measurements to produce an aligned set of points~\cite{Rusinkiewicz:2002,Weise:2009}. Such approaches, however, are limited to scanning objects and scenes of small spatial extent. To extend the scope of point-based methods, Henry et al.~\shortcite{Henry:2012} use surfel primitives, each of which stores a location, a surface orientation, a patch size, and color. More recently, Keller et al.~\shortcite{Keller2013} fuse depth information at the level of points. In recent years, volumetric data structures holding a (truncated) signed distance field (TSDF) representation of scene surfaces~\cite{Curless:1996} have become a common choice for scanning objects and indoor environments. 

KinectFusion~\cite{Newcombe:2011} is the first real-time dense volumetric scanning system that uses a regular TSDF voxel grid of a fixed size to fuse depth measurements. Follow-up works have used octrees~\cite{Zeng:2013} or voxel hashing~\cite{Niessner:2013} to efficiently scale TSDF-based scanning and fusion to larger scenes. In order to minimize camera drift, several methods fuse the incoming data in fragments first and then perform a global optimization~\cite{Zhou:2013,zhou:2013iccv,Choi:2015}. Recently, the BundleFusion system~\cite{Dai:2017} first uses sparse RGB features to achieve a coarse global alignment, and then refines it utilizing geometric and photometric measures. At the core of such approaches, however, is the volumetric data fusion step which assumes that the scanned environment consists of relatively large objects with \emph{closed surfaces}, i.e., objects that have a well-defined inside and outside (at least locally). Thus, such methods are unsuitable for reconstructing thin structures. In contrast, we propose to fuse the sensor measurements at the level of underlying curve structures or skeletons, which are concurrently discovered from the data itself.

Some RGBD methods use the commonly higher pixel resolution of the RGB channel to upsample the depth channel resolution, e.g., by joint filtering~\cite{Kopf:2007:JBU,Richardt:2012:RGBD}, or via shading-based refinement on depth images~\cite{Wu:2014:RSR} or TSDF volumes~\cite{Zollhofer:2015:SRV}. Unfortunately, these methods suffer from the ambiguity between geometry and color edges, and none of them recovers geometry information of thin structures that is already missing in raw depth images or TSDF volumes.

\paragraph*{Reconstructing thin structures} Several methods~\cite{Ummenhofer2013,Savinov_2016_CVPR,tubiblio89320} have recently been proposed to relax the \emph{closed surface} assumption of the volumetric fusion techniques to enable reconstruction of {\em thin surfaces}. While these methods show impressive results, we are addressing a different class of scenes featuring structures that are fundamentally one dimensional and lack sufficient surface detail.


In the specific case of reconstructing objects with similar delicate structures, Li et al.~\shortcite{Li:2010} introduce the deformable model \emph{arterial snakes}. Their method, however, works on high-quality dense 3D scans whereas the input to our method comes from commodity depth sensors and thus has more noise and missing data. Huang et al.~\shortcite{Huang:2013} extract skeletons from point cloud data which can be used to fit generalized cylinders for reconstruction purposes~\cite{Yin:2014}. However, as we show in our evaluations, due to alignment inaccuracies and drift, directly extracting skeletons from merged depth data of all the input frames does not yield compelling results.

An alternative approach to reconstructing thin structures is to use image input. Tabb et al.~\shortcite{Tabb2013} reconstruct thin structures from multiple image silhouettes by fusing information in a volumetric grid using a probabilistic approach \revision{(see Section~\ref{sec:results} for a comparison)}.  Li et al. ~\shortcite{li2018reconstructing} propose a method to solve this problem by leveraging spatial curves generated from image edges, however the quality of the result declines under complex occlusions. Y\"{u}cer et al.~\shortcite{Yucer:2016} present a method that explores local gradient information in captured dense light fields to segment out thin structures. In a follow-up work~\cite{Yucer20163dv}, they combine gradient information with photo-consistency measurements to compute a per-view depth map that can represent thin structures. These depth maps are aggregated using a voxel grid of fixed size, before voxel carving and Poisson surface reconstruction are applied~\cite{Kazhdan:2006}.  In contrast, we propose to directly use extracted 3D skeletons as a fusion primitive to successfully aggregate information from noisy depth and color images captured by commodity depth sensors.

Instead of matching standard point features, several multi-view stereo methods match higher order primitives such as lines~\cite{JainCVPR2010,hofer2014improving} or  curves~\cite{Xiao:05,Fabbri:2010,Rao:2012,Nurutdinova:2015,UsumezbasFK16}. These approaches, however, produce a reconstruction in the form of individual line or curve segments which possibly suffer from noise and gaps. Several methods address this limitation by reconstructing continuous curve paths using different priors. Martin et al.~\shortcite{Martin:2014} reconstruct thin tubular structures such as cables from a dense set of images using physics-based simulation of rods to improve accuracy. This method assumes that 2D cable crossings can easily be recovered from the the images and disambiguated in 3D with an occupancy grid. In contrast, the typical objects we reconstruct often lack a surface, which would make such a grid very noisy. Delmas et al.~\shortcite{delmas2015} reconstruct curvilinear navigating devices (e.g., guide wires, catheters) from two fluoroscopic views as a single continuous path. The more recent work of Liu et al.~\shortcite{Liu:2017} reconstructs objects that are composed of wires utilizing smoothness and simplicity priors. They assume images with relatively clean background (i.e., do not deal with unknown background subtraction) and cannot handle objects with many complex junctions as shown in Section~\ref{sec:results}.

%% file: overview.tex
\section{Overview}

\begin{figure*}[t!]
    \centering
    \includegraphics[width=\textwidth]{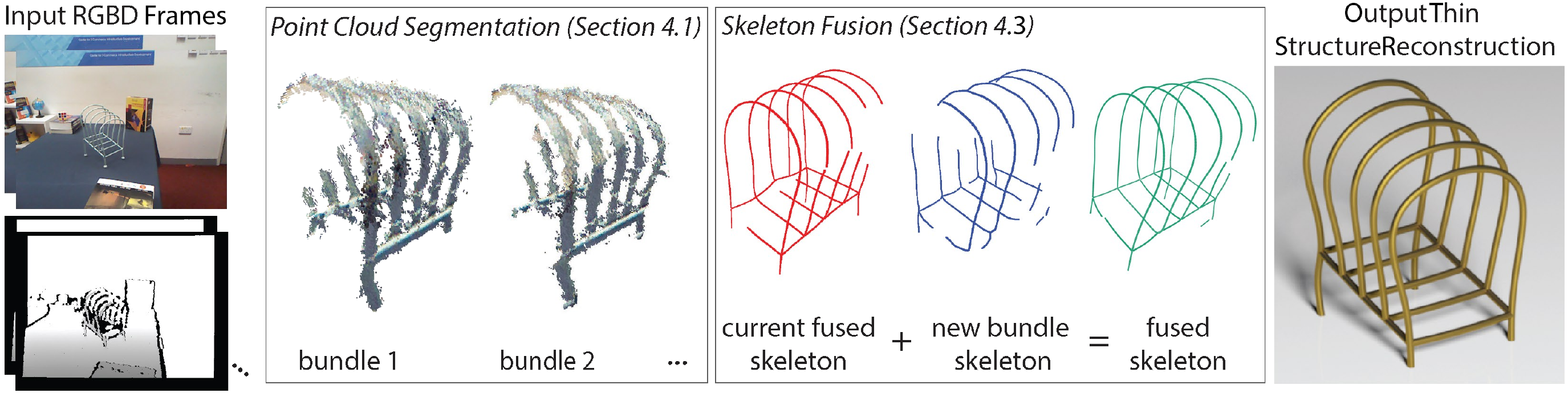}
    \caption{Given a sequence of RGBD images of a thin structure, we organize the input into {\em bundles}, each consisting of $30$ frames. For each bundle, our method first identifies the 3D point samples that belong to the thin structure (Section~\ref{sec:pointSelection}) and extracts the \Lone\ skeleton for each bundle. We then perform a novel skeleton fusion step to iteratively merge the new incoming bundle skeleton to the current partial skeleton (Section~\ref{sec:skeletonFusion}). The final fused skeleton provides a faithful reconstruction of the thin structure.}
    \label{fig:overview}
\end{figure*}

Our goal is to reconstruct a coherent thin structure, denoted $\mathcal{T}$, using a handheld RGBD camera producing a streaming sequence of RGBD frames. Please refer to Figure~\ref{fig:overview} for illustration. 
Note that thin structures smaller than $2$~mm in diameter are usually too thin to be scanned by a consumer-grade RGBD camera, such as Kinect V1, while tubular objects of $10$~mm or larger in diameter can usually be scanned using the same camera and reconstructed in reasonable quality using a conventional pipeline, such as KinectFusion. 

We precompute relative camera poses between the consecutive frames using the ORB-SLAM system~\cite{orbslam} \revision{(in Section~\ref{sec:results}, we evaluate performance using other calibration methods)}. The output of our method is a {\em skeleton} defined as a network of curves in 3D that are the central curves of the thin structure $\mathcal{T}$ along with a radius function. 


\revision{There are two essential challenges in reconstructing accurate skeletons of thin structures from noisy RGBD measurements. First, the {\bf data segmentation} issue: it is difficult to distinguish point samples belonging to the thin structure  $\mathcal{T}$ from those belonging to the other scene parts (i.e., background). Specifically, the depth channel is noisy and unreliable, and it is non-trivial to differentiate thin structures from texture edges in the RGB image alone (see Figure~\ref{fig:RGBD_motivation}b). 
Second, the {\bf data consolidation} issue: in a single depth image, the part of the point cloud that represents the  thin structure  $\mathcal{T}$ is typically very noisy, sparse and incomplete, thus preventing reliable skeleton extraction. On the other hand,  the attempt to extract the curve skeleton from the aligned depth scans of all the frames also fails (see Figure~\ref{fig:RGBD_motivation}c) as the thin-structure gets smeared out in the accumulated point cloud due to error accumulation from camera drift. 
}

\revision{To address the {\bf data segmentation} issue, we separate the point samples belonging to the thin structures $\mathcal{T}$  from those belonging to large objects in the background by first running the traditional volumetric fusion with a regular TSDF voxel grid~\cite{Curless:1996}, henceforth VolumeFusion. (Note that VolumeFusion is given access to registration information obtained using ORB SLAM during camera localization.) The rationale is that the volumetric fusion methods perform quite well in reconstructing objects with relatively large extended surfaces, thus providing a useful reference for distinguishing the thin structure from the background objects.  Then, we propose a simple and effective segmentation method based on {\color{black}topological operators} that combines observations in the RGB images, raw depth maps, and depth renderings of the volumetric fusion result (Section~\ref{sec:pointSelection}). Specifically, a comparison of depth renderings of the fusion result with raw depth maps provides strong cues to remove the depth samples that belong to the background scene (i.e., extended surfaces). Verifying this cue with 2D image space curves in the RGB channel leads to reliable segmentation. }

\revision{To address the {\bf data consolidation} issue, we partition the entire input image frame sequence into {\em bundles}, each being a group of a fixed number of consecutive frames, and merge the segmented point samples from all the depth frames of each bundle into a point set, called a {\emph{bundle difference set}}. Each bundle typically contains $k$ merged frames (in our tests, $k=30$) and provides a good balance between density and drift. The bundle difference sets are dense enough for skeleton extraction, and yet do not suffer from camera drift as severely as merging all the raw depth frames.}

\revision{Once all the bundle difference sets are available, we perform \Lone\ skeleton extraction~\cite{Huang:2013} in each bundle to obtain the skeleton for the part of $\mathcal{T}$ that is observable in the bundle, which we call the \emph{bundle skeleton}. Note that such skeleton curves offer a natural and novel geometric representation as a fusion primitive for reconstructing thin structures. Finally, we propose a novel fusion procedure to topologically aggregate the bundle skeletons into a final 3D skeleton structure (Section~\ref{sec:skeletonFusion}) by aligning and merging skeleton segments. }

%% file: method.tex
\section{\curveF\ Method}
\label{sec:approach} 

Given a sequence of RGBD images providing depth samples from the entire scene, first we  segment  the measurements belonging to the (unknown) thin structure $\mathcal{T}$.

\paragraph*{Assumptions}
We make the following  assumptions about the thin structures being scanned: (i)~Thin structures are sufficiently separated ($\sim$2-5cm depending on depth) from other large surfaces in the background such that the limited depth resolution of commodity sensors is sufficient to capture samples at different depths; (ii)~Thin structures neither (fully) absorb RGBD camera's IR light (e.g., black surface color) nor are highly reflective, which would altogether preclude measuring their depth; (iii)~Thin structures can be distinguished from the background in at least a few RGB frames in each bundle to help utilize image cues; and (iv)~thin structures are in the range $2$~mm to $10$~mm in diameter. \revision{Here, the lower bound of $2$~mm is imposed by the scanning limit of commodity hand-held RGBD scanners, while thin structures of diameter larger than $10$~mm can be reconstructed with reasonable quality using the conventional KinectFusion pipeline.}

\begin{figure}[t!]
    \centering
    \includegraphics[width=\columnwidth]{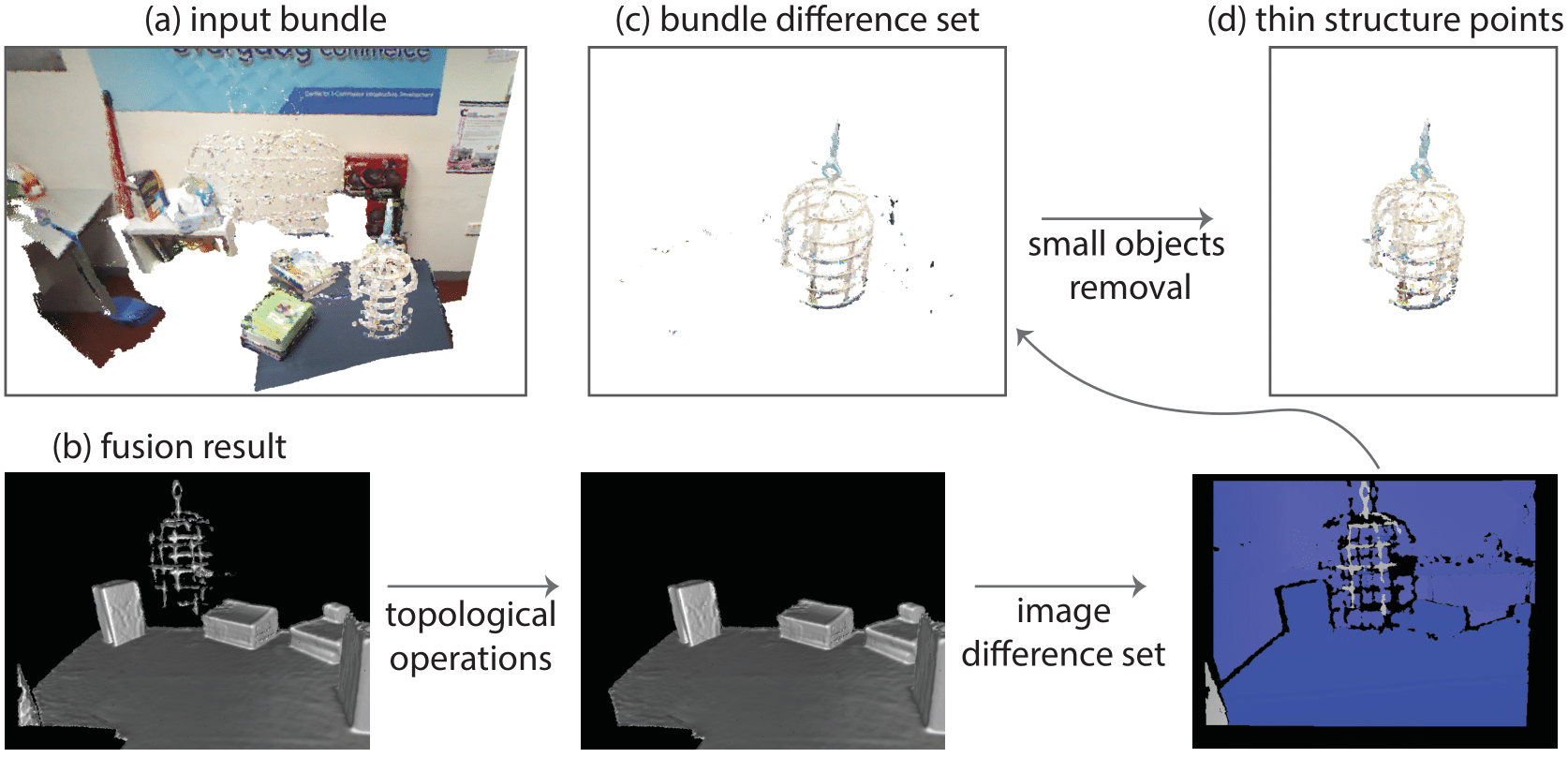}
    \caption{To detect a thin structure $\mathcal{T}$, we first perform topological operations on the depth renderings of the VolumeFusion result and compare them with the input depth images. This produces a set of points in each depth image that potentially belong to $\mathcal{T}$, called the \emph{image difference set}. By combining the image difference sets in a bundle, we obtain the \emph{bundle difference set}. Once isolated ghost point clusters are removed, we obtain point samples that belong to $\mathcal{T}$ only. }
    \label{fig:ptSegmentation}
\end{figure}

\subsection{Point Cloud Segmentation}
\label{sec:pointSelection}

Under the above assumptions, we observe that 3D points captured in each depth image can be classified as one of the following types: (i)~belonging to thin structures to be detected and extracted; (ii)~belonging to large background surfaces to be removed; and (iii)~small and isolated erroneous `ghost' points to be removed. Hence, we propose an automatic segmentation method that detects and removes points of the second and third types (see Figure~\ref{fig:ptSegmentation}). 

\paragraph*{Removing large surfaces}
The result of applying VolumeFusion on the raw depth maps, henceforth called {\em fusion result} (see Figure~\ref{fig:ptSegmentation}b), is a volumetric TSDF model of the entire scene that represents large objects and extended surfaces well, but misses the thin structure $\mathcal{T}$ partially or entirely. This suggests that comparing each depth image against the aligned fusion result helps to identify and remove points belonging to large objects if they are present in both. However, depth points actually belonging to $\mathcal{T}$ may be erroneously removed here, if,  against all odds, the thin structure was partially captured in the fusion result. 



To address this, we first identify parts of $\mathcal{T}$ reconstructed in the fusion result before comparing it to raw depth maps. Specifically, we  use ray-casting to produce a depth rendering of the fusion result from the associated camera pose of each input frame. 
Then, we detect edges in each of these depth renderings by analyzing depth discontinuity, and apply topological operations, namely erosion and dilation, to conservatively remove the  corresponding thin parts that were captured in the VolumeFusion result. 
We set the distance used for both erosion and dilation to $2$~cm. 
Difference of these resulting depth renderings from the corresponding original depth maps yields a point cloud called {\em image difference set} for each input frame as shown in Figure~\ref{fig:ptSegmentation}.

The image difference sets of all frames in the same bundle are aligned to the first frame using the camera poses provided by ORB SLAM. Thus, each bundle is associated with the camera parameters of its first frame.
This aligned and merged point set is called the {\em bundle difference set}, from which large objects have been removed (see Figure~\ref{fig:ptSegmentation}c). 
We resample this merged point set, which is typically non-uniform and dense, using a 3D regular grid for more efficient skeleton extraction at later stages. 

\paragraph*{Removing `ghost' points}
As explained before, the bundle difference set may potentially contain unwanted small and isolated `ghost' objects, i.e., thin and short point cloud segments that do not correspond to any physical 3D scene element. Such ghost objects often arise due to  measurement noise from commodity RGBD cameras. 

\revision{Our goal is to disambiguate these ghost objects from the actual thin structures in each bundle different set. Specifically, we first run the DBSCAN algorithm~\cite{Ester:1996} to segment each bundle difference set into distinct point clusters. We then remove clusters smaller than an empirically determined size threshold (fewer than 10 points) and isolated floating clusters (>10 cm away from any other cluster). The remaining possible thin and long clusters are further removed if they are not \emph{verified} by the RGB channel. Specifically, we first represent all the clusters in a voxel grid. Then we detect curve edges in each RGB frame of the bundle and lift points sampled on these edges to the voxel grid of the clusters. Here, the corresponding raw depth measurements is used for this lifting operation. Then any cluster for which less than 20\% of its voxels receive a lifted curve sample is removed. These operations leave us with a set of points, for each bundle, that only belong to the thin structure, as illustrated in Figure~\ref{fig:ptSegmentation}.}

\revision{We note that due to noise in image space curve edges and raw depth measurement, simply lifting all the detected RGB edges to 3D would produce an extremely noisy point set not suitable for extracting thin structures (see Figure~\ref{fig:RGBD_motivation}). Nonetheless, the RGB image provides sufficient cues to validate candidate thin structure clusters produced by our proposed segmentation method.}

\subsection{Skeleton Extraction}
\label{sec:skeletonExtraction}

Having identified the point clusters in each bundle that belong to the thin structure $\mathcal{T}$, we next extract the central skeleton of each cluster of the bundle by computing the \Lone\ medial axis of the cluster~\cite{Huang:2013}, which is called the {\em bundle skeleton}.

\begin{figure}[h!]
    \centering
    \includegraphics[width=\columnwidth]{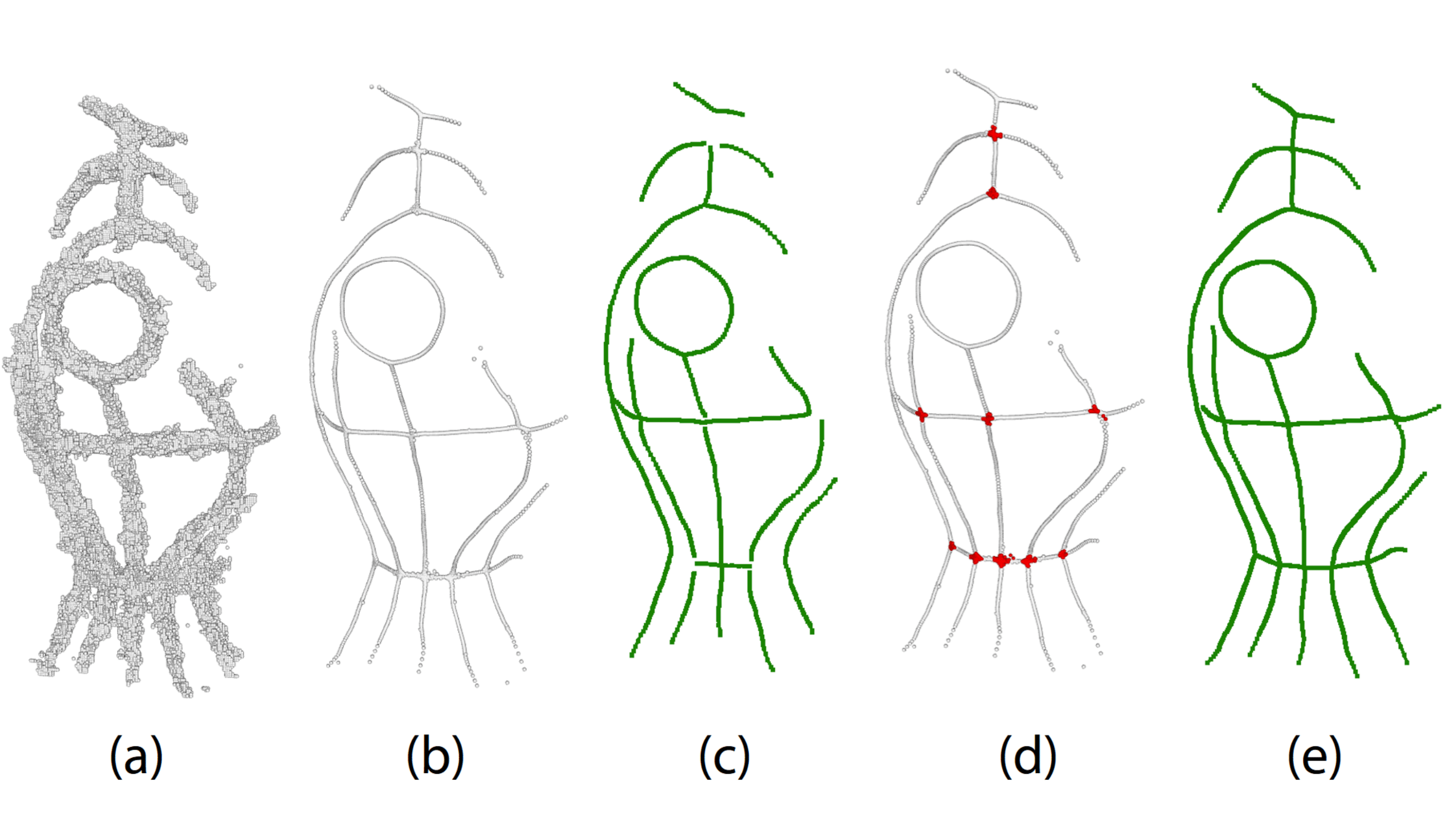}
    \caption{Starting from an input point cloud~(a), the original method of Huang et al.~\shortcite{Huang:2013} leads to spurious junctions (b, c). We explicitly detect junctions~(d),  which leads to better quality \Lone\ skeletal curves~(e).  }
    \label{fig:l1_skeleton}
\end{figure}

Since the \Lone\ axis method~\cite{Huang:2013} was originally designed for  objects (such as aircraft and human hands) that are quite different from our thin tubular structures,  modifications were needed to make it suitable for our setting. In our implementation, we follow the basic framework of \Lone\ contraction with the initial radius set to 15~mm and the maximum radius set to 25~mm based on the typical width of the point cloud of the wire objects we expect to reconstruct. Figure~\ref{fig:l1_skeleton}(b) shows an \Lone\ contraction result of a point set (Figure ~\ref{fig:l1_skeleton}(a)). The original \Lone\ method uses a rather sophisticated strategy to obtain skeleton branches. However, their method often fails in our setting, especially for wire objects with closely adjacent joints, as shown in Figure~\ref{fig:l1_skeleton}(c). We address this problem by identifying the junctions explicitly. After the convergence of contraction, each cluster of the remaining non-branch points is treated as a junction, as shown in Figure~\ref{fig:l1_skeleton}(d). Finally, the skeleton branches are connected by junction points. Empirically, this leads to better quality of \Lone\ skeleton curves for thin objects, as shown in Figure~\ref{fig:l1_skeleton}(e).

\subsection{Skeleton Fusion}
\label{sec:skeletonFusion}
Our goal is to obtain the complete skeleton $\mathcal{K}$ of  the thin structure $\mathcal{T}$ by fusing the bundle skeletons $\mathcal{S}_i$ of the bundles $\mathcal{B}_i$, $i=1,2,\dots, n$. We achieve this goal by performing iterative skeleton fusion. We maintain a partial skeleton $\mathcal{K}_i$ obtained by fusing the bundle skeletons $\mathcal{S}_j$ of all the bundles $\mathcal{B}_j$, $j=1,2,\dots, i$, processed thus far. Prior to this fusion process, we first bring each bundle skeleton to the reference frame of the first bundle using the camera parameters associated to each bundle.
We represent each individual bundle skeleton as well as the partial skeleton $\mathcal{K}_i$ as a graph of connected curves in 3D -- each edge is a curve segment and each node is a junction where a number of curve segments are joined. Initially, $\mathcal{K}_1$ is set to be $\mathcal{S}_1$.


\begin{figure}[h!]
    \centering
    \includegraphics[width = \columnwidth]{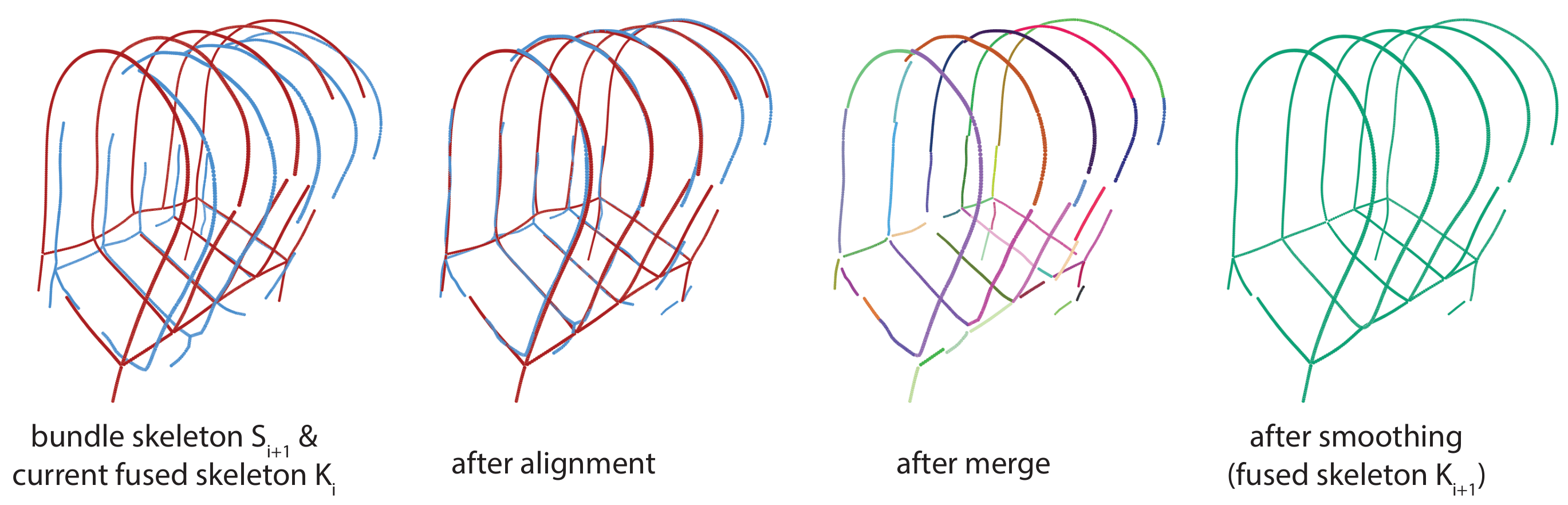}
    \caption{Given the current partial skeleton $K_i$ (red) and the new bundle skeleton $S_{i+1}$ (blue), we show the different steps of the fusion algorithm to generate $K_{i+1}$ (green).}
    \label{fig:fusion}
\end{figure}

Given a partial skeleton $\mathcal{K}_i$, $i\geq 1$, we now describe the details of a single iteration of the skeleton fusion process where the bundle skeleton $\mathcal{S}_{i+1}$ of the next bundle $B_{i+1}$ is fused with the current partial skeleton $\mathcal{K}_i$ to generate $\mathcal{K}_{i+1}$. 
%
As shown in Figure~\ref{fig:fusion}, each skeleton fusion step comprises three subsequent sub-tasks, namely (i)~\emph{alignment}; (ii)~\emph{merging}; and (iii)~\emph{smoothing}.

\paragraph*{(i) Alignment} 
While registering all bundle skeletons to the reference frame of the first one brings them sufficiently closer, there still remain significant misalignments due to camera drift. 
To rectify this, we further bring  $\mathcal{K}_i$ closer to $\mathcal{S}_{i+1}$ by using iterative closest point~(ICP)~\cite{Besl:1992} to perform rigid alignment between densely and uniformly distributed sample points on the curves of $\mathcal{S}_{i+1}$ and $\mathcal{K}_i$.
We do not consider the alignment of the junctions in this step, since the closest point pairs sampled on the curves provide enough constraints to determine the rigid motion.

\begin{figure}[h!]
    \centering
    \includegraphics[width = \columnwidth]{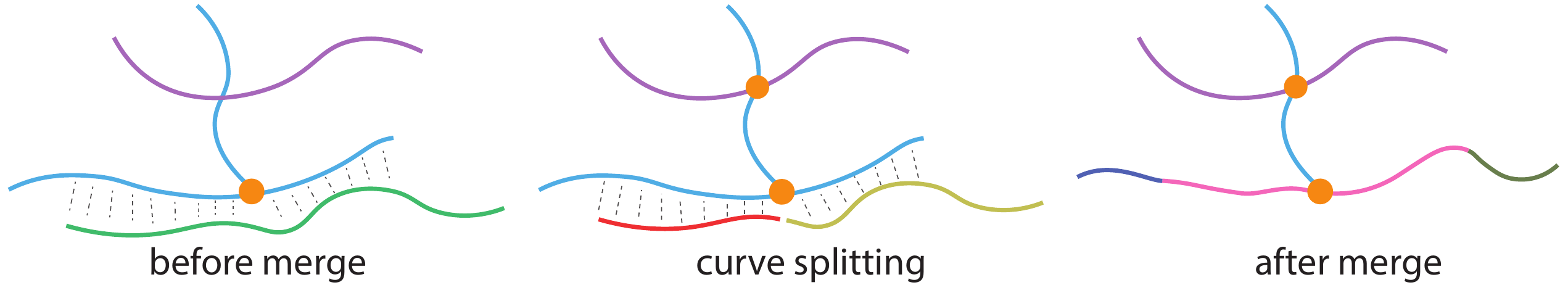}
    \caption{During merging a segment is split when it crosses a junction in the merged region.}
    \label{fig:merge}
\end{figure}

\paragraph*{(ii) Merging} Once $\mathcal{S}_{i+1}$ and $\mathcal{K}_i$ are rigidly aligned, we first detect overlapping curve segments, i.e., curves which are close to each other within a specified distance threshold (set to be 1~cm in our experiments). We merge the overlapping regions of such curve segments by weighted averaging, with the weight $i/(i+1)$ for $\mathcal{K}_i$ and $1/(i+1)$ for $\mathcal{S}_{i+1}$, while keeping the original curve segments for the non-overlapping areas. If a junction point exists in the overlapping region, we split the curve segment that crosses the junction and merge the resulting segments separately (see Figure~\ref{fig:merge}).

Once the curve segments are merged, we identify pairs of junctions of $\mathcal{S}_{i+1}$ and $\mathcal{K}_i$ that are close to each other within a specified distance threshold (set as 1~cm) and merge them with a similar weighted averaging. Any curve segment incident to one of the merged junctions now becomes incident to the new junction.

After merging, if any two curve segments intersect (within a proximity threshold) at an internal point, we create a new junction at this intersection and split each of the segments accordingly. Finally, we remove any curve segment shorter than a threshold (2~mm) to trim off possible spurious parts in reconstruction. The outcome of this merging step is a new network $\mathcal{K}_{i+1}$ of connected curves that inherits the junctions and curves from the skeletons $\mathcal{S}_{i+1}$ and $\mathcal{K}_i$, as well as the new junctions created by curve segment intersections.

\begin{figure*}[h!]
    \centering
    \includegraphics[width =\textwidth]{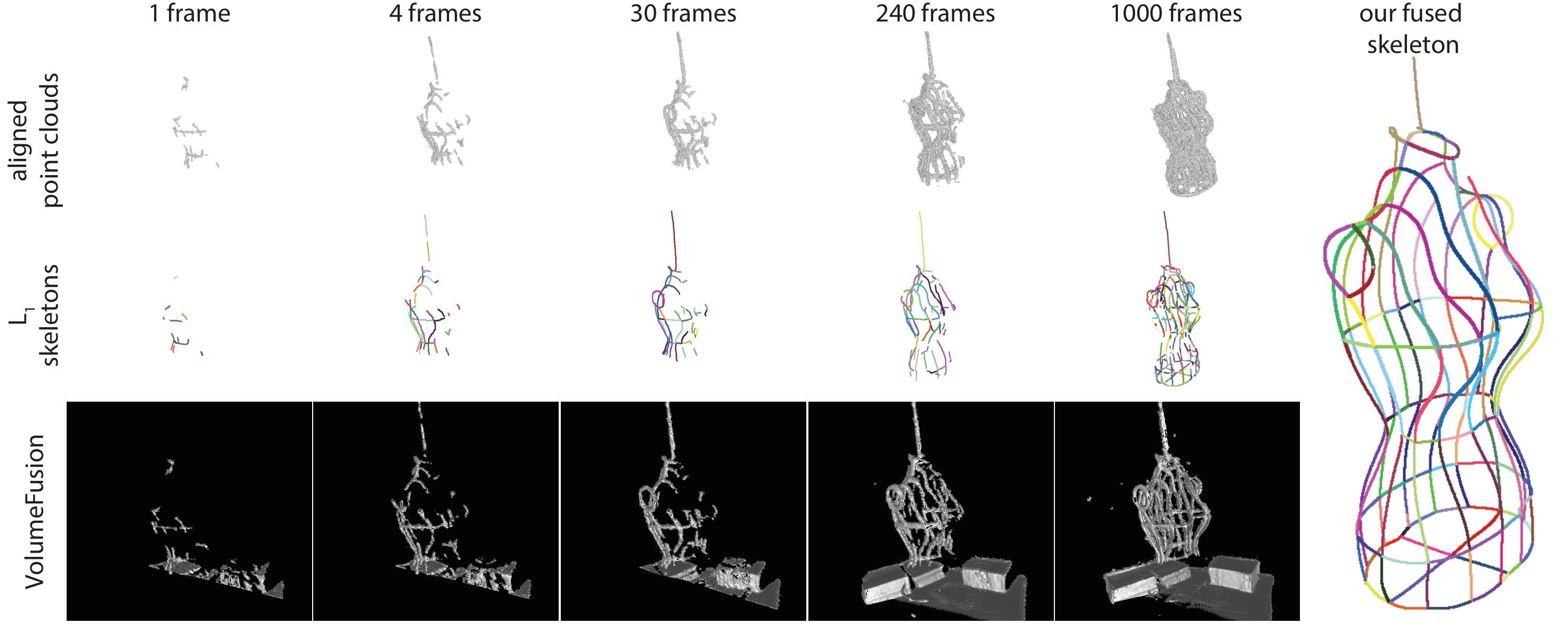}
    \caption{For different choices of bundle size, we show the aligned point clouds and extracted \Lone\ skeletons. We observe that a bundle of $30$ frames provides a good balance between model completeness and structural clarity. Our skeleton fusion algorithm merges per-bundle \Lone\ skeletons to yield a high quality reconstruction. VolumeFusion fails to produce satisfactory reconstruction from any accumulated point cloud, regardless of the number of frames used.}
    \label{fig:bundle}
\end{figure*}

\begin{figure*}[t!]
    \centering
    \includegraphics[width=.97\textwidth]{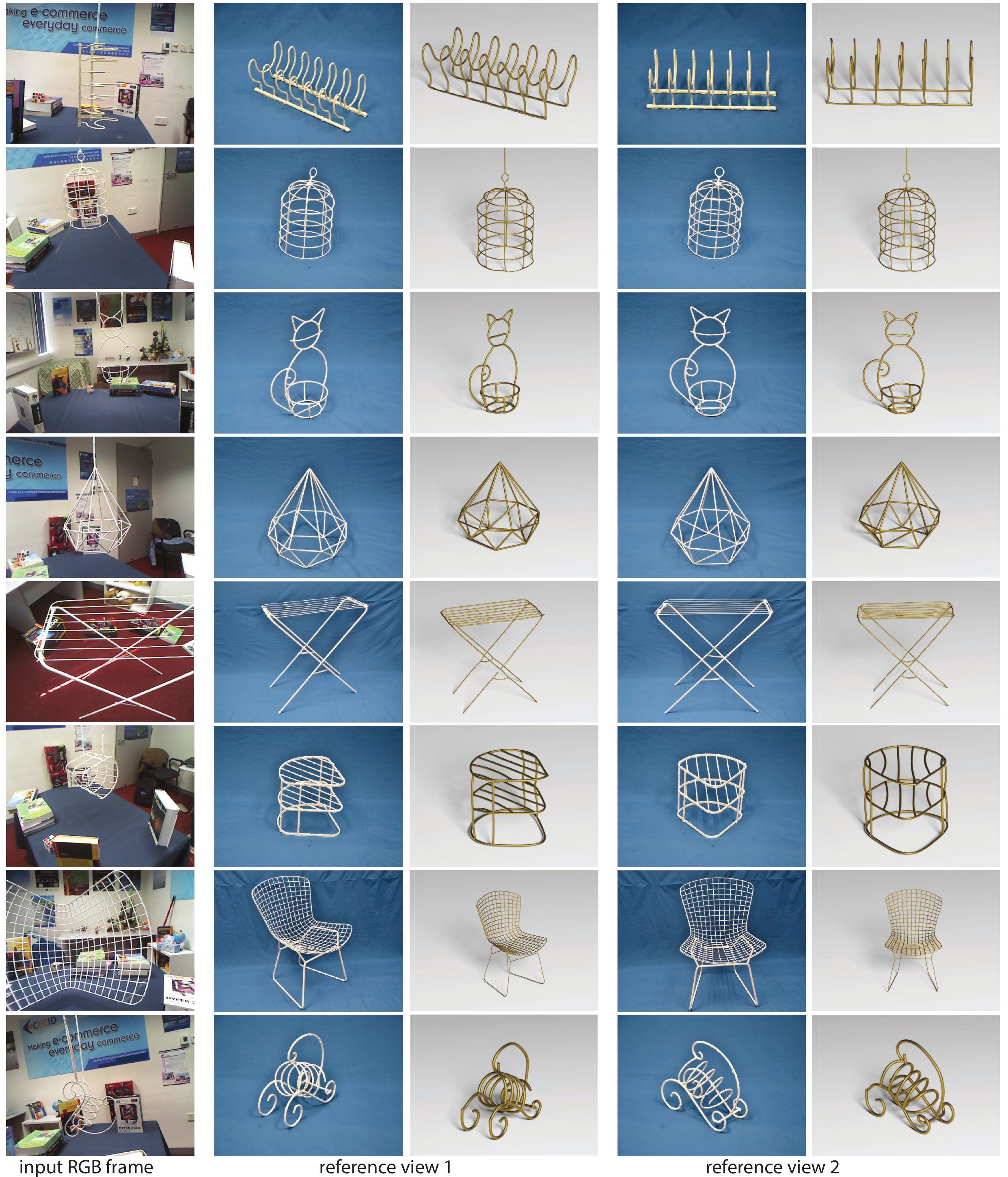}
    \caption{For each example, we show one of the input RGB frames and our reconstruction from two reference views. We show images of the 3D models from the same reference view on a blue background to better convey their geometry (see also video).}
    \label{fig:results}
\end{figure*}

\paragraph*{(iii) Smoothing} The previous step can introduce artifacts in the form of jagged junctions produced by the weighted averaging of overlapping curve segments. To remove these, we perform skeleton smoothing via optimization-based curve fitting. Specifically, we represent each curve of $\mathcal{K}_{i+1}$ as a polyline whose vertex positions $\{v_0, \dots, v_m\}$ are uniformly sampled along the curve to ensure a distance of 1 mm between consecutive vertices. The curve fitting optimization computes the new vertex positions by minimizing the following objective function:
\begin{equation}
\sum_{k=0}^{m} \| v_k - v^0_k\|^2 + \lambda  \sum_{k=1}^{m-1} \| v_{k-1} - 2v_k + v_{k+1}\|^2,
\label{smooth}
\end{equation}
where $v^0_k$ denotes the position of vertex $k$ before smoothing. The first term is a data fitting term which advocates for maintaining the original vertex positions, the second term enforces smoothness in the direction of consecutive polyline segments, and $\lambda$   denotes the relative weighting between these terms ($\lambda=60$ in our tests). 

In order to smooth all the curves in the skeleton simultaneously, we treat each junction as a shared vertex for all of its incident curve segments. Hence, we formulate and minimize a global objective function that sums up the energy given in Equation~\ref{smooth} for each of the individual curves in the network $\mathcal{K}_{i+1}$. This is a linear least squares problem that is solved efficiently by solving a linear system of equations.

\paragraph*{Wire radius estimation} We assume that the thin structure $\mathcal{T}$ is made of tubular surfaces represented by skeleton curves with some constant radius. After the skeleton curve $\mathcal{K}$ of $\mathcal{T}$ is extracted as described in the previous step, we estimate the radius of $\mathcal{T}$ using its image in the RGB frames.
Specifically, we sample a set of points from the skeleton $\mathcal{K}_i$ of each bundle and then project these points from 3D to the RGB images associated with the bundle. Note that edge extraction has been applied to each RGB frame to label the images of the wire object as a strip region. We then inspect a $16\times16$ box (in pixel) centered at the projection of each sample point. The box is accepted for further processing only if there is exactly one strip region in the box, and discarded otherwise. This ensures that boxes with spurious edges due to cluttered background, where radius estimation would be unreliable, are discarded. The width of the strip region is measured for each pixel on the central axis of the strip contained in the box, and then all these width estimates are averaged to produce the radius estimate from this box. 

All accepted boxes (from all sample points on all bundle skeletons) produce a radius estimate in pixels, and these estimates are filtered and combined to produce the final radius measurement as follows. 
When a wire object is composed of multiple parts of different 
radii, as it is the case for some of the objects we processed, our method still produces a single radius value. In this case, the radius measurements from different boxes are clustered 
based on a histogram of the measurements and the most prominent value is 
chosen as the final radius measurement. The resulting 
measurement is a reasonable approximation for the 
majority of the parts having the same radius, even if radii differ across the object.
Finally, the radius in pixels is converted into a metric 3D measurement.

%% file: results.tex
\section{Results}
\label{sec:results}

We tested \curveF\ with a range of real world structures of varying complexity. All our real scenes were recorded with a Kinect V1 sensor. Since the accuracy of the factory calibration between RGB channel and depth channel (captured by two separate sensors) is insufficient for our purpose, we calibrate the two channels ourselves using a checkerboard visible in both IR and color. We show a wide range of examples in Figure~\ref{fig:results}. For each of these, we show an example RGB image from the original scanning sequence, as well as additional representative pictures to better convey the expected geometry. We refer to the accompanying video for close-ups. 

Most of the models shown in this paper were reconstructed from an RGBD sequence of several hundred frames with a bundle size of $30$ frames. Some complex modes took a longer sequence -- for example, the cloth hanger shown in the teaser was scanned in 1,136 frames. Our examples have skeleton structures that contain from 14 to 336 junctions. Our algorithm successfully detects and classifies 95\% of these junctions and provides a faithful 3D reconstruction.

\paragraph*{Effect of bundle size} In Figure~\ref{fig:bundle}, we demonstrate how we determine the appropriate number of frames in a bundle. The first row shows the accumulated point clouds (only points identified to belong to the thin structure are merged) of different numbers of depth frames for a wire model. Clearly, when too few frames are grouped together, e.g. $k=1$ or $4$, the merged point cloud is too sparse and leads to a broken partial skeleton (see the second row of Figure~\ref{fig:bundle}). On the other hand, when too many frames are grouped together, e.g. $k=240$ or $1000$, the merged point cloud becomes more dense but too blurry (due to camera drift) for extracting reliable curve skeletons. Hence, we propose to group the input sequence of frames into bundles of $k$ frames each in such a way that the merged point cloud of each bundle attains a balance between model completeness and structure clarity. The appropriate number of frames $k$ in each bundle typically lies in a range from 10 to 80, depending on model variety and variations of scanning operation. Empirically, we find that $k=30$ works well for most of the models we have tested. As a comparison, the third row of Figure~\ref{fig:bundle} shows that VolumeFusion fails to produce satisfactory reconstruction from any accumulated point cloud, regardless of the number of frames used.

\begin{figure}[t!]
    \centering
    \includegraphics[width = \columnwidth]{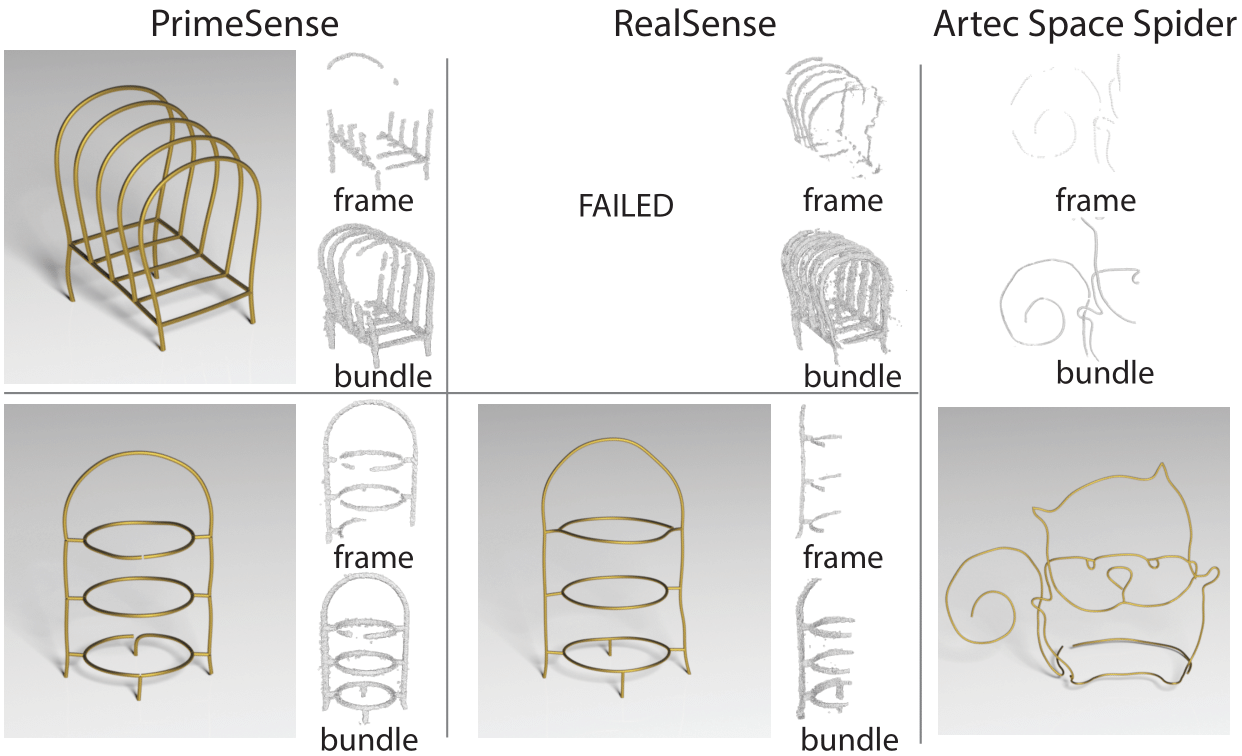}
    \caption{While we can reconstruct both the breadholder (5~mm in diameter) and the rack (8~mm in diameter) models with PrimeSense, RealSense data is much more noisy and we can only use it to reconstruct the thicker rack model. Artec Spider Scan is a high end scanner and thus enables scanning model with wire diameter 1.5~mm. In addition to final reconstructions, we also provide a sample of a single depth frame and a bundle.}
    \label{fig:diff_sensors}
\end{figure}

\paragraph*{Effect of different sensors} In addition to Kinect V1 sensor used for the results presented, we also evaluated four other scanners: PrimeSense, RealSense SR300, Kinect V2, and Artec Space Spider. 
Since PrimeSense's working mechanism is similar to that of Kinect V1, it produces depth scanning of comparable quality. RealSense SR300 produces more noisy depth frames and hence are not suitable for reconstructing very thin structures with our method. However, it can be used for reconstructing thicker structures. Figure~\ref{fig:diff_sensors} shows the results of using these two sensors to scan two models, the breadholder (diameter=5~mm) and the rack (diameter=8~mm).  Note that the breadholder cannot be reconstructed with RealSense. Finally, being a time-of-flight depth sensor, Kinect V2 yields very noisy measurements for thin structures and cannot be used for producing any reasonable reconstruction results. We have also tested Artec Space Spider, a high-end professional grade depth scanner, which is capable of scanning wires as thin as 1.5~mm in diameter. \curveF\ successfully reconstructs using such data as shown in Figure~\ref{fig:diff_sensors}. In summary, as the accuracy and the depth resolution of the scanners increase, the thickness of the wire structures that can be successfully reconstructed from the corresponding depth scans decreases. 

\begin{figure}[h!]
    \centering
    \includegraphics[width = \columnwidth]{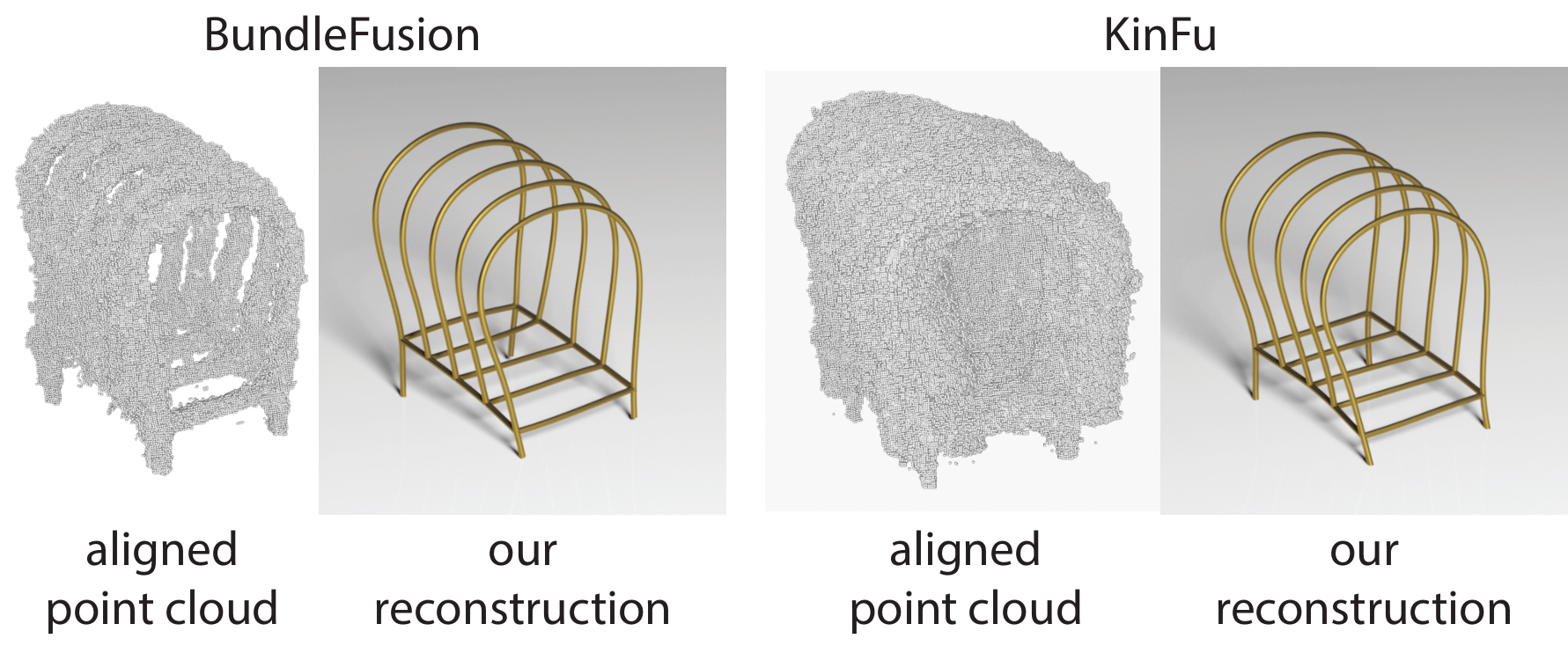}
    \caption{We show two reconstructions of the breadholder with two different camera calibrations provided by BundleFusion and KinFu, respectively. We also show the point cloud in each case obtained by merging all the 399 input depth frames aligned with the given calibration.}
    \label{fig:diff_slams}
\end{figure}

\paragraph*{Effect of different camera calibration} In our pipeline, we used ORB-SLAM, which is an open-source SLAM system using both visual features in RGB frames and geometric features in depth frames. We also tested several other SLAM systems, namely BundleFusion, KinFu~\cite{Kinfu}, and PTAM~\cite{PTAM}. Camera calibration function embedded in BundleFusion performed as well as ORB-SLAM (see Figure~\ref{fig:diff_slams}(a)). 
KinFu is an open source equivalent of KinectFusion that uses geometric features for camera calibration. Given enough background geometric objects, it is able to provide reasonable camera calibration (see Figure~\ref{fig:diff_slams}(b)). 
PTAM, on the other hand, uses only visual features and failed to provide sufficient quality camera calibration to be used by our method. In summary, we conclude it is desirable to employ both visual and geometric features for camera registration, as implemented in ORB-SLAM.

\begin{figure}[h!]
    \centering
    \includegraphics[width = \columnwidth]{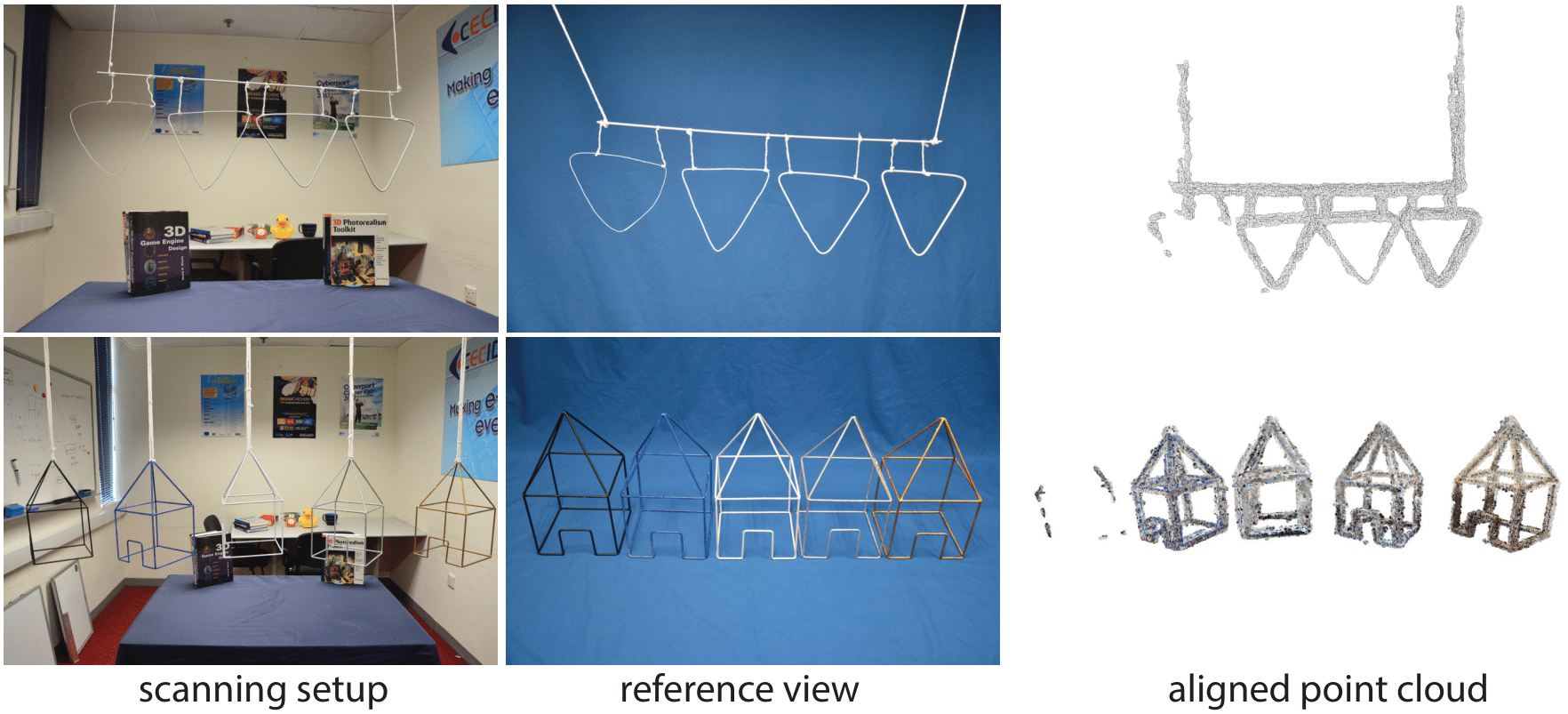}
    \caption{We show the point clouds provided by a depth sensor for models of varying diameter (top row) and color/shininess (bottom row). }
    \label{fig:diff_color_raw_data}
\end{figure}

\paragraph*{Effect of wire thickness and color} Figure~\ref{fig:diff_color_raw_data} shows the point clouds of four triangle wires of different diameters of 1~mm, 2~mm, 3~mm and 4~mm, respectively. We can see that thin structures of diameter smaller than 2~mm cannot be sufficiently scanned by the Kinect V1 for proper reconstruction. Figure~\ref{fig:diff_color_raw_data} also shows the point clouds of another set of similar models in different colors and surface shininess, scanned with Kinect V1. We observe that the leftmost model in black color, though of the same diameter as the others, cannot be scanned using Kinect V1 due to its light-absorbing surface. Note that, the two models on the right have considerable surface shininess but can still be scanned well with Kinect V1.

\paragraph*{Effect of different RGB resolutions} When segmenting the point samples belonging to the thin structure, our method uses RGB frames to perform validation based on image edges. We evaluated the effect of the resolution of RGB frames on reconstruction quality. We considered the two settings of RGB resolution provided in Kinect V1: $480 \times 640$ and $1024 \times 1280$, while fixing the depth image resolution at $480 \times 640$, and observed no discernible difference in the output quality. This is mainly due to the fact that RGB information is used only for verifying the point cluster segmentation suggested by the depth data.

\paragraph*{Effect of inter-wire distance} We tested \curveF\ using a set of parallel straight wires vertically arranged with decreasing spacings. This test served the purpose of determining the lower limit for the intra-wire spacing above which our method can still resolve different tubular structures correctly. Our tests show that the skeleton extraction step works robustly for spacings larger than 17~mm. 

\paragraph*{Comparisons} We compare our approach to 
Volumetric TSDF~\cite{Curless:1996}, KinectFusion~\cite{Newcombe:2011}, and BundleFusion~\cite{Dai:2017} in Figure~\ref{fig:RGBD_motivation}. As we discussed earlier, such approaches that use a fixed size volumetric grid as a fusion primitive either fail to reconstruct the thin structures entirely or provide only a noisy and partial reconstruction. In contrast, our approach detects and accumulates fusion primitives in the form of thin structure skeletons.

\begin{figure}[h!]
    \centering
    \includegraphics[width = \columnwidth]{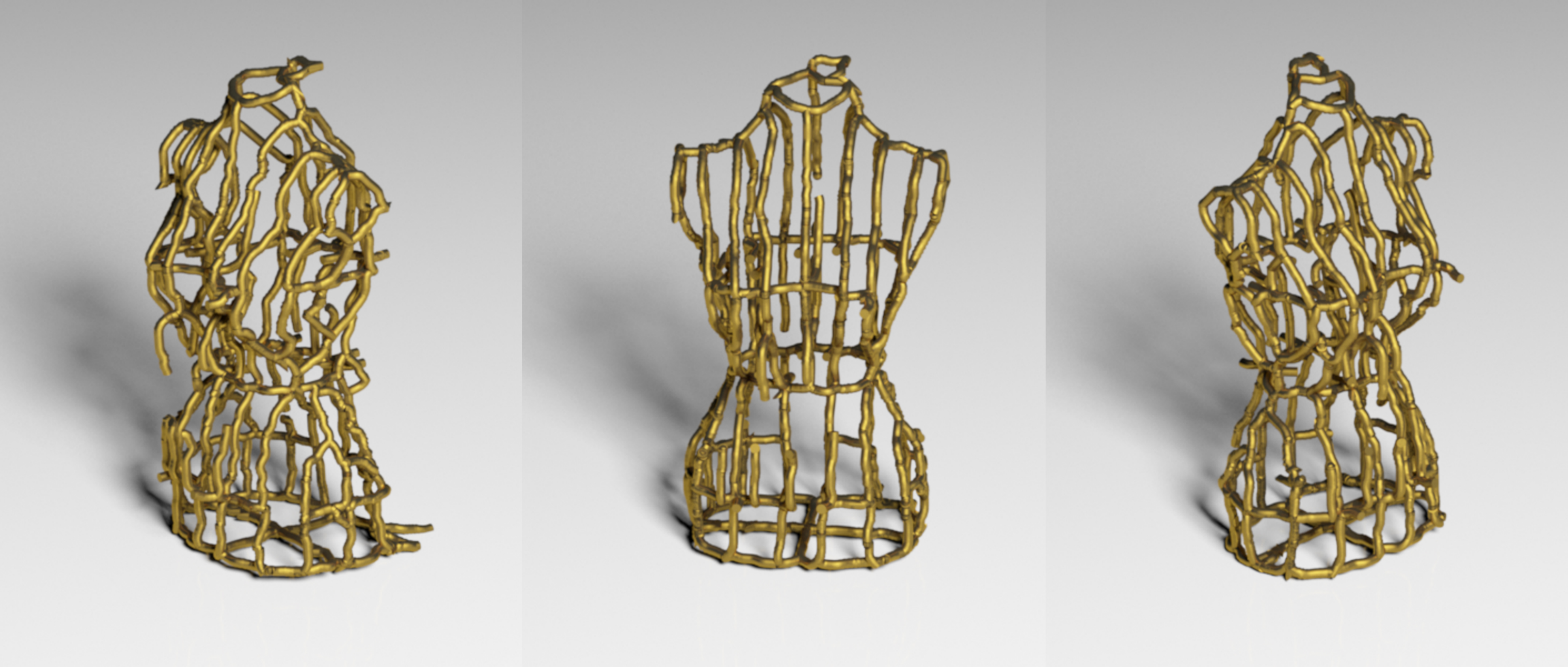}
    \caption{We provide a comparison to the visual-silhouette based method of Tabb et al.~\shortcite{Tabb2013} (compare with Figure~\ref{fig:teaser}).}
    \label{fig:amy}
\end{figure}

\revision{We also provide comparisons with the state-of-the-art visual-silhouette based method of Tabb et al.~\shortcite{Tabb2013}. The heavy studio setup with 40 calibrated cameras used in the original experiment has since been dismantled and has been replaced by a camera installed on a robot arm capable of taking photos of a thin structure from arbitrary viewpoints. With the assistance of the original authors, we tested this new setup together with the original visual silhouette based method for reconstructing the cloth hanger model and show the result in Figure~\ref{fig:amy}. Our result, as can be seen in Figure~\ref{fig:teaser}, provides a smoother and more complete reconstruction while using a much more light-weight setup.}

\begin{figure}[t!]
    \centering
    \includegraphics[width = \columnwidth]{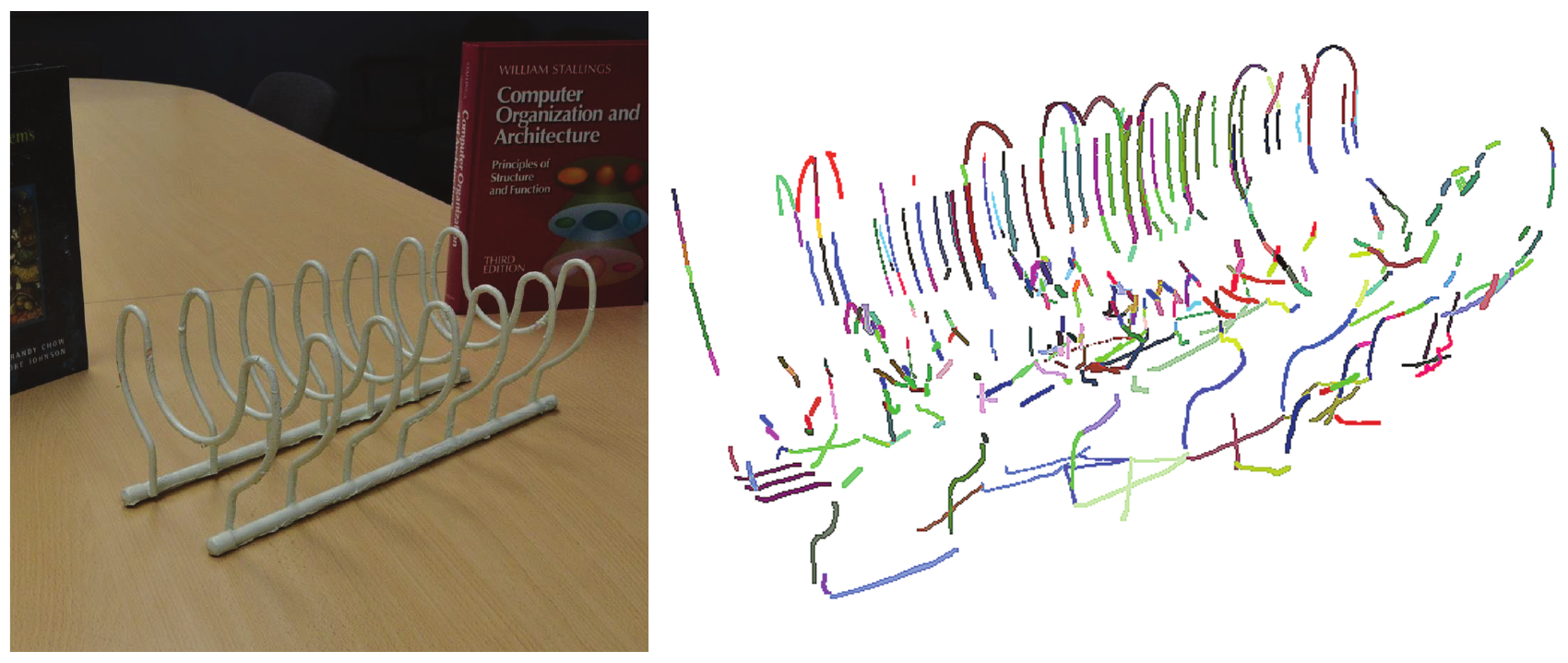}
    \caption{For dense wire structures, assuming access to precise camera calibration and clean background, the image-based reconstruction method of Liu et al.~\shortcite{Liu:2017} results in many spurious 3D curves leading to significant time complexity as it requires to approximate the mTSP problem.}
    \label{fig:sig2017}
\end{figure}

Finally, we perform a comparison to the recent image-based wire reconstruction method of Liu et al.~\shortcite{Liu:2017} (see Figure~\ref{fig:sig2017}). Given $3$ input images of a thin structure with known camera parameters (VisualSFM~\cite{VisualSFM} used to obtain the camera information), this method first detects a set of 2D curves in each image. Then, a large set of 3D candidate curves are generated using epipolar constraints between a pair of images. Candidates which do not receive sufficient support from the third view are then discarded. We observe that in presence of cluttered background and moderately dense wire structures, this image-based reconstruction and verification approach results in many spurious 3D curves which are hard to disambiguate in later stages of the algorithm. Our method, on the other hand, uses the image curves only as a verification cue. We identify potential thin structure point clusters by using the depth information and retain those that also receive image support. Furthermore, Liu et al.~\shortcite{Liu:2017} utilize smoothness and simplicity priors whereas the models we reconstruct are typically composed of many distinct wires that come together at sharp junction points.

\paragraph*{Performance} We measure the execution time of different stages of our method on a machine with Intel(R) Core(TM) i7-6820HK CPU @ 2.70GHz, 16GB memory. While the point cloud segmentation takes about 8 seconds for each bundle, \Lone\ skeleton extraction takes about 12 seconds and remains as a bottleneck. Finally, each iteration of skeleton fusion takes about 1-3 seconds. Please refer to Table \ref{tab:model_inf} for more information.

\begin{table*}
	\centering
	
	\caption
	{
		Statistics and timings corresponding to the results presented in the paper. 
	}  
	
	\resizebox{1\textwidth}{!}{
		
		\begin{tabular}{ | l | c | c | c | c | c | c | c | c |}
			
			\hline
			
			 & \textbf{dish rack} & \textbf{bird cage} & \textbf{cat} & \textbf{decoration} & \textbf{large hanger} & \textbf{rack} & \textbf{chair} & \textbf{wine rack} \\ \hline \hline
			
			\textbf{number of frames}   & 684  & 227& 165& 330 & {261} & {180} & 428& {656}  \\ \hline
			
			\textbf{number of junctions} & 14  & 50& 22& 14&  30& 32& 336& 12 \\ \hline
			
				
			\textbf{thickness of wire}  & {5~mm;  9~mm}  & {4.5~mm} & {4.5~mm; 6~mm} & {5.5~mm} & 4~mm; 12~mm& 5.5~mm; 11~mm& {4.5~mm; 12~mm} & 6~mm  \\ \hline
			\textbf{estimated thickness of wire}  & {5.4~mm}  & {4.8~mm} & {5.6~mm} & {5.1~mm} & 4.5~mm & 5.8~mm& {4.5~mm} & 5.4~mm  \\ \hline
			\tabincell{c}{\textbf{reconstruction time (s)}\\ \textbf{(without L1 extraction)}}    &178	 & 89	 & 37	 & 64	 & 86	 & 55	 & 111	 &  142
			
			\\ \hline
			
		\end{tabular}
		
	}
	
	\label{tab:model_inf}
	
\end{table*}

\paragraph*{Limitations} 
Our method may fail when the assumptions stated in Section~\ref{sec:approach} are violated in the input. 
If a thin structure, or a part of it, is close to a wall or lies on a table (i.e. a resting surface), then in general it cannot be scanned as an object distinct from the resting surface due to the limited depth resolution of the depth sensor. Hence, it will be missing from the reconstruction. 
As another limitation, our method implicitly assumes the RGBD images capture traces of points arising from thin structures, even if partial and noisy. In case of very thin structures below $2$~mm in diameter (e.g., threads or very thin wires), commodity RGBD scanners are unlikely to record any evidence in the depth channel, and hence, \curveF\ will fail to reconstruct such thin structures. \revision{We note that this minimum diameter value can vary depending on the accuracy and the resolution of the sensor as discussed in Section~\ref{sec:results}.} 
For objects with junctions having high valence, our junction merging step can produce imperfect results. For example, a junction of valence 6 may instead be interpreted as two separate junctions of valence 3 each and joined together. The challenge here is that both the depth and color channels are too unreliable to sufficiently disambiguate such situations. A scanner with input of higher accuracy may partially address this issue. Finally, our radius estimation assumes that a thin structure has a constant radius for all its parts. So a better method needs to be devised for radius estimation when reconstructing a thin structure of varying radius.

%% file: conclusion.tex


\section{Conclusion}

We presented \curveF\ as the first algorithm to produce high quality 3D reconstruction of thin filament-like structures from commodity RGBD sequences. We demonstrated that existing state-of-the-art fusion approaches that align and integrate noisy measurements over fixed primitives (e.g., voxels) are unsuited to this task. Instead, as our key contribution, we described how to first reliably extract underlying thin-structured object skeletons from raw RGBD sequences, and then perform surface reconstruction using a data-dependent fusion primitive. We presented several high-quality reconstructions of complex thin-structure objects using our method.

Several future research directions remain to be explored: 

\paragraph*{Towards a real-time system} We would like to speed up our method to work in real-time. Given the current breakdown of timings, the main bottleneck is in detecting \Lone\ axes. One possible direction is to use data-driven learning approaches that infer local properties directly from point sets. 

\begin{figure}[h!]
    \centering
    \includegraphics[width=.9\columnwidth]{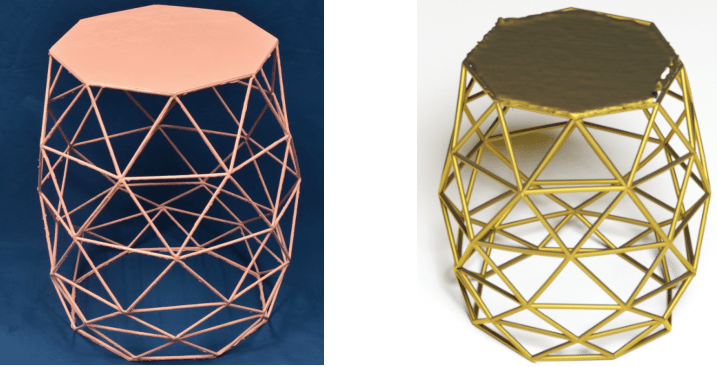}
    \caption{We would like to consider a hybrid fusion algorithm that handles both thin structures and extended surfaces. Right shows a first result where the we manually combined KinectFusion output and \curveF\ results. }
    \label{fig:future}
\end{figure}

\paragraph*{Hybrid structures}
In the current formulation, \curveF\ handles thin structures only. However, many real world objects have a mixture of thin and extended surfaces. In such cases, it would be interesting to design a hybrid approach where voxels are used to recover the surfaces, and extracted skeletons to recover the thin structures (see Figure~\ref{fig:future}). The challenge is to automatically assign a fusion method to each part. 

\paragraph*{Different thickness}
In this work we have assumed that a thin structure has the same radius in all of its parts. We will study how to extend our method to reconstructing thin structures that consist of parts of different radii or have a varying radius, possibly by performing more accurate segmentation of thin structures in the RGB images. 

\paragraph*{Dynamic structures}
Finally, many thin structures, being light-weight, are easily affected by surrounding movements, e.g., plants or hanging lights swaying under wind effects. We would like to extend our method to also capture such dynamic structures borrowing ideas from DynamicFusion~\cite{dynamicFusion15} and variations designed for a comparable surface setting. 